
\input lisboa.sty
\input epsf.sty
\raggedbottom
\footline={\hfill}
\rightline{FTUAM 99-32}
\rightline{hep-ph/9910399}
\rightline{October, 1999}
\bigskip
\hrule height .3mm
\vskip.6cm
\centerline{{\bigfib  Heavy Quarkonium}\footnote*{\petit Lectures
 given at the XVII Autumn School, ``QCD: Perturbative or Nonperturbative", Lisbon, 
September-October 1999.}\phantom{xyasdfrtyuiiop}}
\medskip
\centerrule{.7cm}
\vskip1cm

\setbox9=\vbox{\hsize65mm {\noindent\fib F. J. 
Yndur\'ain} 
\vskip .1cm
\noindent{\addressfont Departamento de F\'{\i}sica Te\'orica, C-XI,\hb
 Universidad Aut\'onoma de Madrid,\hb
 Canto Blanco,\hb
E-28049, Madrid, Spain.}\hb}
\smallskip
\centerline{\box9}
\bigskip
\setbox0=\vbox{\abstracttype{Abstract} Bound states of heavy $\bar{q}q$
 quarks are reviewed within the context of QCD. First of all, we consider 
the calculations which can be performed {\sl ab initio}, 
which includes $\bar{t}t$ with principal quantum number $n$ up to four,
 $\bar{b}b$ states with $n=1$ and (to a lesser extent)
 $\bar{b}b$ with  $n=2$ and 
 $\bar{c}c$ for $n=1$.  Among
 the results, we report a very precise $O(\alpha_s^4)$ evaluation 
of $b,\,c$ quark masses from quarkonium spectrum with a potential to two loops, 
a calculation of the decay $\upsilonv\to e^+e^-$  and a prediction for the 
splitting $\upsilonv\;-\;\eta_b$. We then consider 
how, with the help of reasonable assumptions, one can extend QCD calculations 
to other states of heavy quarks. Finally, a few words are said on 
the treatment of light quark bound states.}
\centerline{\box0}

\pageno=1
\brochureb{\smallsc f. j.  yndur\'ain}{\smallsc heavy quarkonium}{1}
\brochureendcover{Typeset with \physmatex}

\booksection{1. Introduction}
In the present lectures we are going to review some aspects of the  analysis
 of heavy quarkonia, $\bar{t}t$, $\bar{c}c$ and especially $\bar{b}b$ states. 
Before the advent of asymptotic freedom in 1973,  
 hadronic interactions were analyzed with (among other methods) 
the help of the {\sl quark model}, which incorporated a somewhat
 inconsistent set of semiphenomenological rules. 
An important r\^ole was played by bound state calculations in the so-called 
{\sl constituent quark model}, developed in the 
early sixties by, among others,  Morpurgo, Dalitz and collaborators, and 
Oliver, Pene, Reynal and Le~Yaouanc. In this 
model $u,d,s$ quarks 
were given phenomenological masses of $300- 500\,\mev$, and were bound 
by potentials: the harmonic oscillator potential being 
a popular choice because of its simplicity. Quite surprisingly, a large 
number of properties of hadrons could be reproduced in this way.

After the advent of asymptotic freedom, and 
with it of a consistent field theory of strong interactions, it was possible to 
reformulate the quark model in terms of QCD. Thus, De R\'ujula, Georgi and Glashow\ref{1} 
showed that taking into account relativistic corrections and colour algebra 
one could calculate the spectrum of the then known hadrons, including 
in particular such properties as the $N-\Delta$ mass difference, and even the 
$\Sigma^0-\Lambda$ splitting, something that 
had defied previous, non-QCD analyses. They were also able to {\sl predict} 
 qualitative features of the charmonium spectrum.

Nowadays we expect more from QCD, at least for heavy quarks. The reason is that there, 
and to leading order in $\langle v^2\rangle$ 
($\langle v^2\rangle$ the average velocity of the quarks),
 the interaction is equivalent to a {\sl potential}. 
At very short distances this potential has to be of the coulombic type,
$$-\dfrac{C_F\alpha_s}{r}.\equn{(1.1)}$$ 

 Even in the static limit, in QCD, (1.1) will  
be modified by 
radiative corrections; but these should be of the form of a function of $r$. 
In fact, also at long distances one expects, in the nonrelativistic (NR) limit, 
a local potential with the form  
of a function $U(r)$, although it will not be of coulombic type.
 The reason for this is galilean invariance, that will hold in the NR limit. 
By virtue of it, we must have that the derivative of the position ${\bf Q}$, the velocity, 
should be proportional to the momentum:
$$[H,{\bf Q}]=\dfrac{\ii}{\hbar}\dot{\bf Q}=\dfrac{\ii}{\hbar m} {\bf P},$$
with $H$ the hamiltonian. If we define the interaction by 
$H_{\rm int}\equiv H-{\bf P}^2/2m$ and evaluate the commutator above, 
it follows that $[H_{\rm int},{\bf Q}]=0$ and so, because of 
a well-known theorem of von~Neumann, 
 $H_{\rm int}$ must be a function of ${\bf Q}$. A function  
which due to rotational invariance, and at least neglecting spin, may only depend on $r$:
$$H_{\rm int}=U(r).$$

Needless to say, {\sl relativistic} corrections will in general not 
be representable by local, momentum-independent potentials, as is the case even in QED. 
In QCD one encounters QED-like corrections and
 idiosincratic QCD ones, in particular those 
 associated with the complicated structure of the 
 vacuum. Of these the most important are the effects 
 involving the gluon condensate $\langle \alpha_s:G^2:\rangle$,
 first studied in this context by Leutwyler and Voloshin\ref{2}; the 
quark condensate also gives contributions but, for heavy quarkonium, subleading ones.

We will consider in these lectures bound states of heavy quarks (and at the end say a few 
words about light quarks\fnote{Light quark bound states, and gluonic 
bound states, are reviewed in the companion lectures by Yu. Simonov.})
 in {\sl de}creasing order of tractability by rigorous QCD. 
First of all, we will consider situations where one has the inequalities
$$a\ll R;\quad |B_n|\gg \lambdav ;\quad m\gg\lambdav$$
where $a$ is the equivalent of the Bohr radius, and 
$B_n$ are the equivalent of the Balmer energies; $m$ is 
the quark mass, and $\lambdav\simeq 300\,\mev$ is the QCD parameter. 
Under these circumstances, nonperturbative (NP) and confinement effects 
are expected to be small, and so are the radiative and relativistic corrections. 
All of them may therefore be treated as {\sl perturbations} of a nonrelativistic, unconfined 
and leading order (in the QCD coupling $\alpha_s$) systm. This 
will constitute the bulk of these lectures. Then 
we will diverge  from the situations where one can effect 
{\sl ab initio} QCD calculations, relying more and more on 
reasonable (but unproven) assumptions, and on models.

\booksection{2. Heavy quarks at short distances: pure QCD analysis}
For very heavy $\bar{q}q$ bound states  the equivalent of the Bohr radius,
 $a=2/(mC_F\alpha_s)$, is much smaller than the  
confinement radius, $R\sim\Lambdav^{-1}$. So we expect that, for lowest $n$  
 states, with $n$ the principal quantum number, 
 confinement may be neglected, or at least treated as 
a first order perturbation. In this case the quarkonium system is very
 similar to a familiar one, viz., positronium; so that 
methods analogous to those developed 
for the last may be applied also to the study of the 
former. The NR  
 potential may be  obtained 
from {\sl perturbative} QCD; note that, unlike for positronium, 
and because of the zero mass of the gluons, radiative corrections 
are present even in the static limit. 
The leading piece is given by the tree level nonrelativistic
 amplitude and we then include higher effects as perturbations: 
we proceed in steps.

We will also follow the method of {\sl equivalent potentials}, 
advocated for QCD by Gupta and collaborators\ref{3}: 
we find this method the more transparent one. Other, equivalent methods, 
based on the Bethe--Salpeter equation or effective lagrangians may be found in the 
literature\ref{4}. 

In the method of equivalent potentials one profits from the fact that, 
in the NR limit, the potential is given by 
the Fourier transform of the 
scattering amplitude, in the Born approximation:
$$T^{\rm Born}_{NR}({\bf p}\to{\bf p}')=
-\dfrac{1}{4\pi^2}\int\dd^3{\bf r}\,\ee^{\ii {\bf r}({\bf p}-{\bf p}')}V(r),
\equn{(2.1)}$$
where $\bf p$, ${\bf p}'$ are the initial and final momenta in the center of mass reference system. 
On the other hand, $T^{\rm Born}_{NR}$ may be calculated as the nonrelativistic limit of 
the relativistic scattering amplitude:
$$T^{\rm Born}_{NR}\,=\lim_{ m\to\infty}\,\dfrac{1}{4\sqrt{p_{10}p_{20}p'_{10}p'_{20}}}
F(p_1+p_2\to p'_1+p'_2);
$$
formally the NR limit is equivalent to taking the limit of infinite quark masses,
 keeping the three-momenta fixed. One can thus calculate the Born
 approximation to $F$, $F^{\rm Born}$ using 
the familiar Feynman rules (actually at tree level)
 for $\bar{q}q$ scattering, take the NR limit and hence obtain 
$T^{\rm Born}_{NR}$. From it, by inverting (2.1) one finds $V$. 
What is more, we can calculate corrections to $V$ by including corrections 
(in particular, relativistic) to $T^{\rm Born}_{NR}$.
\medskip
\setbox0=\vbox{\hsize 8.5truecm\captiontype\figurasc{\noindent figure 1a.}{A 
ladder of gluon exchanges.
\medskip}}
\setbox1=\vbox{\hsize 10truecm\hfil\epsfxsize=8truecm\epsfbox{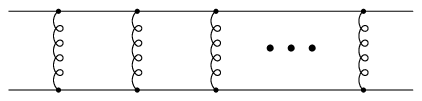}\hfil}
\centerline{\tightboxit{\box1}}
\medskip
\centerline{\box0}

A few words will now be said about the connection of the method 
with Feynman graph calculations; 
more details may be found in \chap~6 of ref.~5, which we roughly follow 
in these notes. Working in the strict  
nonrelativistic limit to avoid inessential complications, it is 
easy to check that solving the 
Schr\"odinger equation with the 
coulombic potential of \equn{(1.1)} is equivalent 
to summing an infinite ladder of exchange graphs (\fig~1{\sc a}): 
indeed, both methods yield the same S matrix. 
Including radiative corrections then alters the potential, as will be shown below. 
The corresponding Schr\"odinger equation will be equivalent to dressing the
 kernel $F_N$ associated 
with the calculated radiative corrections to $N$ loops 
with ladders, corresponding to 
the coulombic wave functions, as shown graphically in 
\fig~1{\sc b} where the kernel would be given, e.g. to two loops, 
by graphs depicted in \figs~2~3 below. Thus the 
method of potentials is  equivalent to a particular
 arrangement of the summation of perturbation theory.
\topinsert{
\setbox0=\vbox{\hsize 8.5truecm\captiontype\figurasc{\noindent figure 1b.}{Kernel,
 dressed with infinite sums of ladders.
\medskip}}
\setbox1=\vbox{\hsize 10truecm\hfil\epsfxsize=8truecm\epsfbox{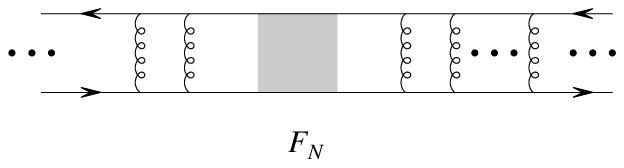}\hfil}
\centerline{\tightboxit{\box1}}
\medskip
\centerline{\box0}
}\endinsert

\booksubsection{2.1. Step 1: coulombic Schr\"odinger equation}
According to what we have said, we proceed in steps. 
In this first step we consider a nonrelativistic tree-level kernel (\fig~2).
Here 
the scattering amplitude is such that it generates a coulombic potential. 
So we get the Schr\"odinger equation, for the energies and 
wave functions, 
$$\eqalign{H^{(0)}&\Psiv_n^{(0)}= E_n^{(0)}\Psiv_n^{(0)}\cr
H^{(0)}=&2m-\dfrac{1}{m}\lap+V^{(0)}(r),\quad V^{(0)}(r)=-\dfrac{C_F\alpha_s}{r};\cr}$$
note that for quarkonium the 
reduced mass is $m/2$. The quantities $m,\;\alpha_s$ however 
are as yet undefined; only 
when including radiative corrections will we be able to give them a precise meaning. 
This equation, 
formally identical to that of an hydrogen-like atom,
 can be solved exactly, and will be our starting point in the calculations.

\medskip
\setbox0=\vbox{\hsize 2.8truecm
\captiontype\figurasc{\noindent figure 2.}{One- gluon exchange.
\vskip2cm
\phantom{x}}}
\setbox1=\vbox{\hsize 9truecm\hfil\epsfxsize=7.2truecm\epsfbox{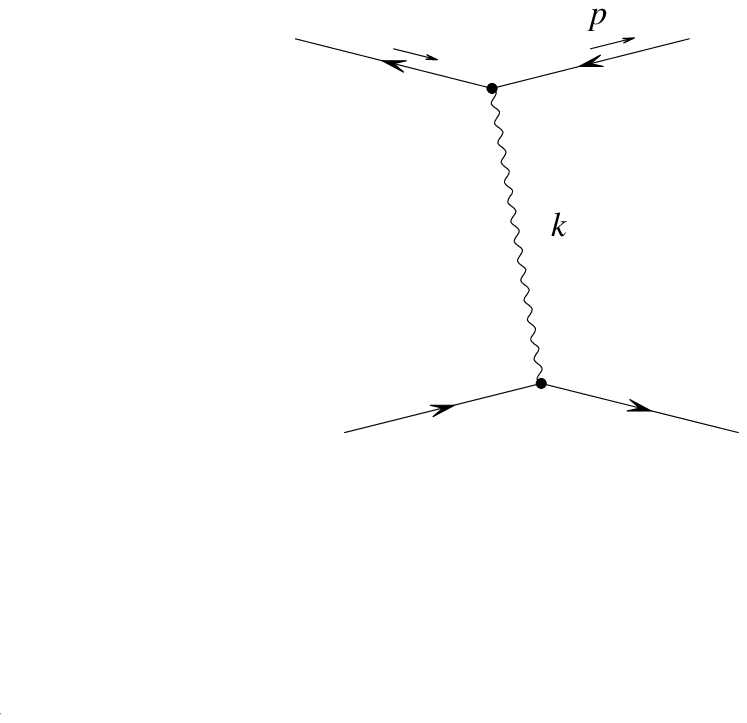}\hfil}
\line{\tightboxit{\box1}\hfil\box0}

\booksubsection{2.2. Step 2: relativistic corrections}
The relativistic corrections are identical to those for 
positronium, known since ancient times. They are found 
by considering still the tree level scattering 
amplitude corresponding to the diagram of \fig~2, but keeping now the terms 
of order $1/c^2$, $c$ the speed of light; 
the details of the derivation may be found in textbooks
 on relativistic quantum mechanics\ref{6}.
 Adding also the 
correction to the kinetic energies, 
$$\sqrt{m^2-\lap}\simeq m-\lap/2m-\lap^2/8m^3$$ 
one finds the hamiltonian,
$$H_{\hbox{``tree"}}=H^{(0)}+V^{(0)}_{\rm rel}\equn{(2.2a)}$$
where the superscript zero in $V^{(0)}$ 
indicates that the potential is still   
obtained from a tree level (zero loop) amplitude. The relativistic corrections, 
which 
 are to be treated as first order perturbations to
 the unperturbed equation (2.1),  read
$$V^{(0)}_{\rm rel}=V^{(0)}_{\rm si;\,rel}+
V^{(0)}_{\rm tens}+V^{(0)}_{LS}+V^{(0)}_{\rm hf}.
\equn{(2.2b)}$$
The various pieces, spin-independent (in 
which we also include the 
correction to the 
kinetic energy), 
tensor, $LS$ and hyperfine, are 
$$\eqalign{V^{(0)}_{\rm si;\,rel}=
&\,-\dfrac{1}{4m^3}\lap^2+\dfrac{C_F\alpha_s}{m^2}\,\dfrac{1}{r}\lap,\cr
V^{(0)}_{\rm tens}=&\,\dfrac{C_F\alpha_s}{4m^2}\,\dfrac{1}{r^3}S_{12},\cr
V^{(0)}_{LS}=&\dfrac{3C_F\alpha_s}{2m^2}\,\dfrac{1}{r^3}{\bf L}{\bf S},\cr
V^{(0)}_{\rm hf}=&\,\dfrac{4\pi C_F\alpha_s}{3m^2}\,{\bf S}^2\delta({\bf r}).\cr}
\equn{(2.2c)}$$
Here $\bf L$ is the orbital angular momentum operator, $\bf S$ the total spin 
operator, and $S_{12}$ 
the tensor operator:
$${\bf L}=-\ii{\bf r}\times\nabla,\quad {\bf S}=\dfrac{\ybf{\sigma}_1+\ybf{\sigma}_2}{2},\quad
S_{12}=2\sum_{ij}\left(\dfrac{2r_i r_j}{r^2}-\delta_{ij}\right)S_iS_j.$$
The wave functions are assumed to have spinor components, 
and the Pauli matrices $\ybf{\sigma}_a$ act on spinor $\chi(\lambda_a)$, $a=1,\,2$.

\booksubsection{2.3. Step 3: radiative corrections}
Before discussing the radiative corrections, a matter has 
to be settled first, which is that of the meaning of the mass in the 
Schr\"odinger equation. We have defined the 
potential by assuming that it vanishes at infinity; otherwise, 
we have the ambiguity of an arbitrary constant.
Thus, we must interpret the mass as the mass at long distances, i.~e., on the mass shell. 
Now, both these requirements are not rigorously valid 
since quarks are confined in a region of radius $R\sim \lambdav^{-1}$; 
but we can work with them to the extent that $R$ is much larger than
 the region where the movement of the quarks takes place, $a=2/mC_F\alpha_s$.

With this requirement we define $m$, called the mass shell, or pole 
mass, to be  such that, {\sl in perturbation theory},
$$S_{\rm p.t.}(\slash{p}=m)^{-1}=0.\equn{(2.3)}$$
The relation of this mass with the \msbar\ mass, $\bar{m}$ was found by 
Coquereaux and Tarrach\ref{7} to one loop and by Gray et al.\ref{8} to two loops. 
After correcting a misprint of the last reference one finds
$$m=\bar{m}(\bar{m}^2)\left\{1+\dfrac{C_F\alpha_s(m^2)}{\pi}
+\left(K-2C_F\right)\left[\dfrac{\alpha_s(m^2}{\pi}\right]^2\right\},
\equn{(2.4a)}$$
where, denoting by $n_f$  the number of 
quark flavours with mass less than or equal to $m$,
$$\eqalign{K=&\tfrac{1}{9}\pi^2\log2+\tfrac{7}{18}\pi^2
-\tfrac{1}{6}\zeta(3)+\tfrac{3673}{288}-\left(\tfrac{1}{18}\pi^2+
\tfrac{71}{144}\right)n_f
+\sum_{i=1}^{n_f-1}\deltav\left(\dfrac{m_i}{m}\right),\cr
\Deltav(\rho)=&\tfrac{4}{3}\left\{\tfrac{1}{8}\pi^2\rho-\tfrac{3}{4}\rho^2
+\cdots.\right\}.\cr}\equn{(2.4b)}$$
\smallskip
\setbox0=\vbox{\hsize 7truecm\captiontype\figurasc{\noindent figure 3.}{Some 
radiative corrections.
\medskip}}
\setbox1=\vbox{\hsize 10truecm\hfil \epsfxsize=8truecm\epsfbox{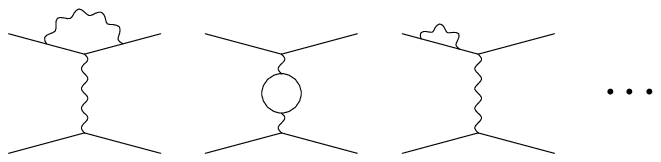}\hfil}
\centerline{\tightboxit{\box1}}
\smallskip
\centerline{\box0}
\medskip

All other quantities, however, are renormalized in the \msbar\ scheme. 
So, $\alpha_s(\mu^2)$ will be the \msbar\ coupling at the scale 
$\mu^2$, etc.
These radiative corrections (some of which are shown in \fig~3)
 have been evaluated by a number of people. Those 
to the spin-independent part of the potential, in the 
strict static approximation, were first calculated by 
Fischler and Billoire\ref{9} to one loop and by Peter
 and Schr\"oder\ref{10} (who checked and corrected a mistake 
of Peter's calculation) to two loops. 
Because of the zero mass of the gluons, corrections nonanalytic in the average 
velocity occur. These depend on 
$|v|$ and are of a size comparable to that of 
the two loop static corrections; note that, in the coulombic approximation, 
$\langle|v|\rangle=C_F\alpha_s$. They were calculated, together with 
 $O(v^2)$ radiative corrections to one loop, in ref.~11.
 Spin-dependent one loop corrections were evaluated in refs.~3, 12.

We discuss in some detail the spin-independent 
part of the spectrum. To take into account all terms giving 
corrections of $O(\alpha_s^4)$ to the energy spectrum one writes the
 Hamiltonian as
$$H=\widetilde{H}^{(0)}+H_1\equn{(2.5a)}$$
where $\widetilde{H}^{(0)}$ may, and will, be solved exactly and contains all the coulombic 
pieces of the interaction:
$$\eqalign{\widetilde{H}^{(0)}=&2m+
\dfrac{-1}{m}\lap-\dfrac{C_F\widetilde{\alpha}_s(\mu^2)}{r};\cr
\widetilde{\alpha}_s(\mu^2)=&
\alpha_s(\mu^2)
\left\{1+\left(a_1+\dfrac{\gammae\beta_0}{2}\right)\dfrac{\alpha_s(\mu^2)}{\pi}\right.\cr
+&\left[\gammae\left(a_1\beta_0+\dfrac{\beta_1}{8}\right)\right.
\left.\left.+\left(\dfrac{\pi^2}{12}+\gammae^2\right)\dfrac{\beta_0}{4}+
a_2\right]\dfrac{\alpha_s}{\pi^2}\right\}.\cr}\equn{(2.5b)}$$
$H_1$ is to be considered as a {\sl perturbation}. Its form is,
$$H_1=V_{\rm s.i.;\,rel}+V^{(L)}_1+V^{(L)}_2+V^{(LL)}+V_{\rm s.rel}+V_{\rm hf},\equn{(2.5c)}$$

$$\eqalign{V_{\rm s.i.;\,rel}=&
\dfrac{-1}{4m^3}\lap^2+\dfrac{C_F\alpha_s}{m^2r}\lap,\cr
V^{(L)}_1=&
\dfrac{-C_F\alpha_s(\mu^2)^2}{\pi}\,\dfrac{\beta_0}{2}\dfrac{\log r\mu}{r},\cr
V^{(L)}_2=&
\dfrac{-C_F\alpha_s^3}{\pi^2}\,
\left(a_1\beta_0+\dfrac{\beta_1}{8}+\dfrac{\gammae\beta_0^2}{2}\right)\dfrac{\log r\mu}{r},\cr
V^{(LL)}=&
\dfrac{-C_F\beta_0^2\alpha_s^3}{4\pi^2}\,\dfrac{\log^2 r\mu}{r},\cr
V_{\rm s.rel}=&
\dfrac{C_Fb_1\alpha_s^2}{2mr^2},\cr
V_{\rm hf}=&
\dfrac{4\pi C_F\alpha_s}{3m^2}s(s+1)\delta({\bf r}).\cr}
\equn{(2.5d)}$$
$V_{\rm s.i.;\,rel}$ is the spin-independent piece of $V_{\rm rel}$ in \equn{(2.2c)}; 
$V_{\rm s.rel}$ is a one loop velocity-dependent correction and 
all other ones are one and two loop static corrections 
with the exception of the last term, 
representing hyperfine splitting.  Although we are considering the 
spin-independent interaction, to the precision we are working one needs to differentiate 
between the masses of vector and pseudoscalar states, hence the presence of this 
piece. 
 $a_1$ was calculated in ref.~9, $a_2$ in ref.~10 and $b_1$ and many of the rest 
of the terms in ref.~11; all these constants are given in the Appendix. 
Note that, to the precision we are working, 
$\alpha_s$ has to be evaluated to three loops
$$\eqalign{\alpha_s(Q^2)=&\dfrac{4\pi}{\beta_0L}\left\{1-\dfrac{\beta_1\log L}{\beta_0^2L}+
\dfrac{\beta_1^2\log^2L-\beta_1^2\log L+
\beta_2\beta_0-\beta_1^2}{\beta_0^4L^2}\right\},
\cr
L=&\log Q^2/\Lambdav^2.\cr}$$

Some bookkeeping is necessary to identify 
which terms to include for a given 
order of accuracy in the calculation, e.g. 
in the evaluation of the energy levels. 
The pieces given here will provide a calculation accurate to 
order $\alpha_s^4$. 
All terms in $H_1$ are to be treated as first order 
perturbations of $\widetilde{H}^{(0)}$, except for
the term $V^{(L)}_1$, which has to be evaluated to second order. Thus it 
produces, in addition to the first order contribution,
$$\eqalign{\delta^{(1)}_{V^{(L)}_1}E_{10}=
-m\dfrac{\beta_0C_F^2\alpha^2_s(\mu^2)\widetilde{\alpha}_s(\mu^2)}{4\pi}
\left(\log\dfrac{a}{2}+1-\gammae\right),\cr}\equn{(2.6a)}$$
 the second-order energy shift, for the ground state,\ref{13}
$$\eqalign{\delta^{(2)}_{V^{(L)}_1}E_{10}=
-m\dfrac{\beta_0C^2_F\alpha_s^4}{4\pi^2}
\left[N_0\log^2\dfrac{a\mu}{2}+
N_1\log\dfrac{a\mu}{2}+N_2\right];\cr}\equn{(2.6b)}$$ 
 the constants $N$ are given 
 in the Appendix.

The first order contributions of the other $V$'s are easily evaluated 
using the formulas of ref.~11. One finds 
$$E^{\rm p.t.}_{nl}=2m-m\dfrac{C_F^2\widetilde{\alpha}_s^2}{4n^2}+\sum_{V}\delta^{(1)}_{V}E_{nl}
+\delta^{(2)}_{V_1^{(L)}}E_{nl}.\equn{(2.7)}$$
The label ``p.t." in $E^{\rm p.t.}_{nl}$ indicates that we have as yet 
only used results deduced from perturbation theory; the full expression would be
$$E_{nl}=E^{\rm p.t.}_{nl}+\delta_{\rm NP}E_{nl},$$
with $\delta_{\rm NP}E_{nl}$ given below, \equn{(2.10)}.
 
The $\delta^{(1)}_{V}E_{nl}$ are, with $a$ as above,
$$\eqalign{\delta^{(1)}_{V_{\rm tree}}E_{nl}=&-\dfrac{2}{n^3\,m^3\,a^4}
\left[\dfrac{1}{2l+1}-\dfrac{3}{8n}\right]+
\dfrac{C_F\alpha_s}{m^2}\,\dfrac{2l+1-4n}{n^4(2l+1)a^3};\cr
\delta^{(1)}_{V_2^{(L)}}E_{nl}=&-\dfrac{C_Fc_2^{(L)}\alpha_s^3}{\pi^2n^2a}\;
\left[\log\dfrac{na\mu}{2}+\psi(n+l+1)\right];\cr
\delta^{(1)}_{V^{(LL)}}E_{nl}=&-\dfrac{C_F\beta_0^2\alpha_s^3}{4\pi^2n^2a}\,
\Big\{\log^2\dfrac{na\mu}{2}+2\psi(n+l+1)\log\dfrac{na\mu}{2}\cr
+&\psi(n+l+1)^2+\psi'(n+l+1)\cr
+&\theta(n-l-2)\dfrac{2\Gammav(n-l)}{\Gammav(n+l+1)}
\sum^{n-l-2}_{j=0}\dfrac{\Gammav(2l+2+j)}{j!(n-l-j-1)^2}\Big\};\cr
\delta^{(1)}_{V_{\rm s.rel}}E_{nl}=&\dfrac{C_Fb_1 \alpha_s^2}{m}\;
\dfrac{1}{n^3(2l+1)a^2}.\cr}\equn{(2.8a)}$$
Here and from now on we have defined $a=2/(C_Fm\widetilde{\alpha}_s)$. 
We recall that constants are collected in the Appendix.
 For the masses of the 
vector states ($\Upsilonv,\,\Upsilonv',\,\Upsilonv'';\;J/\psi,\,\psi',\dots$) 
one has to add the hyperfine shift, at tree level,
$$\delta^{(1)}_{V_{\rm spin}}E_{nl}=\delta_{s1}\delta_{l0}
\dfrac{8C_F\alpha_s}{3n^3m^2a^3}.\equn{(2.8b)}$$
The value of the 
contributions of $V_1^{(L)}$   
were given above, \equn{(2.6)}.

\booksubsection{2.4. Step 4: nonperturbative corrections}
The {\sl leading} nonperturbative (NP) corrections can be shown 
to be those associated with the contribution of the gluon condensate. 
Physically they may be 
understood as follows. We consider that the quarks move in a medium, the QCD vacuum, 
which is full of soft gluons (\fig~4) that we represent by their field strength 
operators, $G^c_{\mu\nu}(x)$. When $a\ll R$, we may assume that the 
confinement size is infinite and, moreover, that one can 
neglect the fluctuations of the $G^c_{\mu\nu}(x)$ in the region of size $a$ 
in which the quarks move. So we  approximate the effect of 
the motion in the gluon soup by 
introducing an interaction, which in the static limit will be of dipole type, 
of the quarks with a constant gluonic field: 
$$H_{NP}=-gr_iG^c_{0i}(0)t^c=-g{\bf r}{\calbf E}^ct^c.$$ 
Because of Lorentz 
invariance of the vacuum we assume that
 $\langle G^c_{\mu\nu}\rangle=0$, but $\langle \alpha_s:G^2:\rangle\neq 0$.  For 
dimensional reasons, this will give the leading NP contribution 
to the spin-independent energy shifts. Applying thus straightforward 
second order perturbation theory we have\fnote{Projectors 
over the subspace orthogonal to  $|\Psiv^{(0)}\rangle$, 
that we do not write explicitely, are understood.}
$$\delta_{\rm NP}E_{nl}=
-\left\langle\Psiv_{nlM}^{(0)},H_{\cal E}
\dfrac{1}{H^{(8)}-E^{(0)}_n}H_{\cal E}\Psiv_{nlM}^{(0)}\right\rangle.
$$
There are a few points  to clarify regarding this equation.
 First of all, we may, since we average over directions, 
neglect the magnetic quantum number $M$. Secondly, we have 
used in the denominator  the {\sl octet} Hamiltonian, 
$$H^{(8)}=-\dfrac{1}{m}\lap+\dfrac{1}{2N_c}\,\dfrac{\alpha_s}{r}.$$
This happens because the perturbed state, 
$H_{\cal E}|\Psiv^{(0)}\rangle=-g{\bf r}{\calbf E}_at^a|\Psiv^{(0)}\rangle,$
is manifestly an octet one, as ${\calbf E}_a$ creates a gluon on 
top of the singlet $|\Psiv^{(0)}\rangle$.
\topinsert{
\setbox9=\vbox{
\medskip
\setbox0=\vbox{\hsize 6truecm\captiontype\figurasc{\noindent figure 4.}{The 
region where the $\bar{q}q$ pair move inside the confinement region.
\vskip1cm
\phantom{x}}}
\setbox1=\vbox{\hsize 6truecm\hfil \epsfxsize 4.truecm\epsfbox{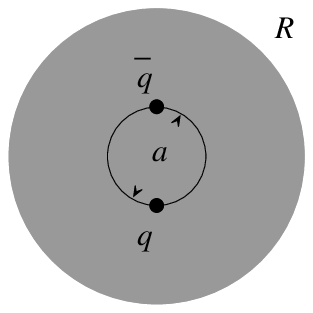}\hfil}
\line{\tightboxit{\box1}\hfil\box0}}
\box9}
\endinsert 

Next, we have to relate expectation 
values of products ${\calbf E}\dots{\calbf  E}$ to the gluon 
condensate. For this, first  write the gluon radiation Hamiltonian as
$$H_{\rm rad}=\dfrac{1}{8\pi}\int\dd^3{\bf r}\,:\calbf{E}^2+\calbf{B}^2:,$$
with sums over omitted colour indices understood. 
Its expectation value in the physical vacuum should vanish, so we conclude
$$\langle{\rm vac}|\int\dd^3{\bf r}:\calbf{E}^2:|{\rm vac}\rangle
=-\langle{\rm vac}|\int\dd^3{\bf r}:\calbf{B}^2:|{\rm vac}\rangle.$$
We assume the field intensities to be constant, so we 
may  replace the integrals by the volume times the integrands at $x=0$.
 Canceling then the volume and recalling that $G^2=-2(\calbf{E}^2-\calbf{B}^2)$, we find
$$\langle{\rm vac}|:\calbf{E}^2:|{\rm vac}\rangle=
-\tfrac{1}{4}\langle{\rm vac}|:G^2(0):|{\rm vac}\rangle.$$
Finally, using Lorentz and colour invariance of the physical vacuum,
$$g^2\langle:\calbf{E}_a^i(0)\calbf{E}_b^j(0):\rangle=
\dfrac{4\pi\alpha_s\delta_{ij}\delta_{ab}}{(D-1)(N_c^2-1)}\langle:\calbf{E}\calbf{E}:\rangle
=-\dfrac{\pi\delta_{ij}\delta_{ab}}{24}\langle\alpha_s G^2\rangle.$$
With this we  get
$$\eqalign{\delta_{\rm NP}E_{nl}=&\,
\left\langle\Psiv_{nl},\left({\bf r}\sum_at^a\calbf{E}_a(0)\right)
\dfrac{1}{H^{(8)}-E^{(0)}_n}H_{\cal E}
\left({\bf r}\sum_bt^b\calbf{E}_b(0)\right)\Psiv_{nl}\right\rangle\cr
=&\,\dfrac{\pi\langle \alpha_s G^2\rangle}{18}\sum_i\left\langle\Psiv_{nl},
r_i\left(-\dfrac{1}{m}\lap+
\dfrac{\alpha_s}{6r}-E_n^{(0)}\right)^{-1}r_i\Psiv_{nl}\right\rangle.\cr}
\equn{(2.9)}$$
To finish the calculation we have to  invert the operator
 $\left(-m^{-1}\lap+\alpha_s/6r\right)$. For the simple case above,
 the method may be found in ref.~2; in more 
complicated situations, cf. ref.~14.
So we finally obtain the nonperturbative pieces of the energy shifts, which are of the 
form\ref{2},
$$\delta_{\rm NP}E_{nl}=
m\dfrac{\pi n^6\epsilon_{nl}\langle \alpha_s:G^2:\rangle}{(mC_F\alpha_s)^4},
\equn{(2.10)}$$
 where the numbers $\epsilon_{nl}$ are of order unity, $\epsilon_{10}\simeq 1.5$.
 The evaluations for 
the spin-dependent shifts may be found in 
 ref.~14 (with a minor correction in ref.~15)
 and the contributions of {\sl higher} order operators
 has been considered in refs.~16,~17. Note that, as already remarked by Leutwyler\ref{2}, 
one cannot derive (2.10) from a local potential; but, 
for the lowest states, the effect may be {\sl approximated}
 by a cubic one,
$$V_{\rm Gluon\,cond.}(r)\sim \Lambdav^4r^3.$$

As promised, the correction (2.10) is relativistic in that it is of order 
$1/m^4$; but the coefficient 
is very large because of the high powers of $n$, $\alpha_s^{-1}$.
 The reason for these powers of $\alpha_s$
 and $n$ can be understood easily. 
Two come from the energy 
denominators, and four from the expectation value
 $\langle r_ir_j\rangle_{nl}\sim n^4/(mC_F\alpha_s)^2$. 
 
The right hand side of (2.10)  grows  as the sixth power of the 
radial quantum number, $n$. It is in fact this very fast 
growth with $n$ that leads to the breakdown 
of the method as soon as $n$ exceeds, or in some cases even equals, the value 2 
for $\bar{c}c$, $\bar{b}b$; only  
for $\bar{t}t$ can one go to $n\sim 5$. 

The NP corrections to the wave function may be obtained 
with  methods similar to those employed to evaluate $\delta_{\rm NP}E$.
 For $n=1,\,l=0$, we have
$$\Psiv_{10}(r)\to\left(1+\delta_{\rm NP}(r)\right)\Psiv_{10}(r),\equn{(2.11a)}$$
where the NP correction is
$$\delta_{\rm NP}(r)=
\left\{\tfrac{2968}{425}-\tfrac{104}{425}\rho^2-\tfrac{52}{1275}\rho^3-\tfrac{1}{225}\rho^4\right\}
\dfrac{\pi\langle\alpha_sG^2\rangle}{m^4(C_F\alpha_s)^6},\quad\rho=\dfrac{2r}{a}.
\equn{(2.11b)}$$
 It turns out that the coefficient of the correction is larger than for the 
energy shifts, both in powers of $\alpha_s^{-1}$ and of $n$:
 the effects of confinement are larger for the 
wave function than for the energy levels.

The contribution of some higher dimensional operators 
has been estimated by Pineda\ref{16}; the corrections due to the finite 
size of the hadron is discussed in ref.~17, and will be briefly reviewed later.

Radiative and nonperturbative corrections to higher excited states 
may likewise be evaluated; as can be calculated the decay rates 
into photons or leptons, e.g., $\upsilonv\to e^+e^-$, $\eta_b\to\gamma\gamma$. 
We will present some of these results below.

\booksection{3.  Results}
Let us summarize the results. The calculation is fully 
justified, in the sense that higher order corrections (both perturbative and NP) 
are smaller than lower order ones for $\bar{b}b$ with $n=1$. The same is partially 
true for the energy levels of the same states with $n=2$ and, for 
$\bar{c}c$, for $n=1$. For the wave functions of $\bar{b}b,\,n\geq 2$ and 
all $\bar{c}c$ states, and for the energy levels with higher 
values of $n$ than the ones reported above, the calculation is 
 meaningless as nominally subleading corrections overwhelm nominally leading ones.

Before presenting the results a few words have to be said 
about the choice of the renormalization 
point, $\mu$. As our \equs~(2.6),~(2.8) show, a {\sl natural} value for this 
parameter is 
$$\mu_0=\dfrac{2}{na}=\dfrac{mC_F\widetilde{\alpha}_s}{n},$$
for states with the principal quantum number $n$, and this will be our choice. For 
states with $n=1$ the results of the 
calculation will turn out to depend  little on the value of 
$\mu$, provided it is reasonably close to $\mu_0$. Higher states are 
another matter; we will discuss our choices when we consider them.

As input parameters we take the recent determinations\ref{18},
$$\Lambdav(n_f=4,\,\hbox{three loops})=0.283\pm0.035\;\gev\;
\left[\;\alpha_s(M_Z^2)\simeq0.117\pm0.024\;\right],
$$ 
and for the gluon condensate, very poorly known, the value
$$\langle\alpha_sG^2\rangle=0.06\pm0.02\;\gev^4.$$
(Note that the slight difference between the results reported below and those 
of previous determinations\ref{11,13} are mostly due to the variation of the 
prefered value of $\lambdav$ from $0.20$ to $0.28$ \gev.)

\booksubsection{3.1.  $n=1$ states}

For $\bar{b}b$ one gets a precise determination of $m_b$, and less so of 
 $\bar{m}_b(\bar{m}_b^2)$ (pole and $\overline{\hbox{MS}}$ masses), a 
reliable prediction for the hyperfine splitting, and reasonable agreement with 
the experimental value of $\Upsilonv~\rightarrow~e^+e^-$; these will be discussed later.
For $\bar{c}c$ a reasonably accurate value  is also obtained 
for $m_c$. For $\bar{b}b$, and with $\lambdav$, $\langle\alpha_s G^2\rangle$ as given before 
and varying $\mu^2$ around $\mu_0^2$ by 25\% 
to estimate the systematic errors 
of the calculation one finds, from the $\Upsilonv$  mass, the 
 quark masses\ref{13}:
$$\eqalign{m_b=&
5.065\pm0.043\,(\Lambdav)\;\mp0.005\,(\langle\alpha_sG^2\rangle)\;\cr
&^{-0.031}_{+0.037} \;(\hbox{vary}\; \mu^2\;{\rm by}\,25\%)
\;\pm 0.006\;({\rm other\; th.\;uncert.})\cr
\bar{m}_b(\bar{m}_b^2)=&
4.455\pm0.012\,(\Lambdav)
\,\mp0.005\,(\langle\alpha_sG^2\rangle)\;\cr
&^{-0.029}_{+0.034} \;
(\hbox{vary}\; \mu^2\;{\rm by}\,25\%)
\;\pm0.006\;({\rm other\; th.\;uncert.}).\cr}
\equn{(3.1)}$$
Note that $m_b$ is correct to $O(\alpha_s^4(\mu_0^2))$, and 
$\bar{m}_b(\bar{m}_b^2)$ to $O(\alpha_s^2(m_b^2))$.  
The piece denoted by the expression ``other th. uncert." in (3.1) 
refers to the error coming from 
higher dimensional operators and some higher order perturbative terms; 
it can be found discussed in refs.~13,~16. It 
 is comfortably 
smaller than the errors due to the uncertainty on $\Lambdav,\;\langle\alpha_sG^2\rangle$.

The values of $\mu^2_0$, $\alpha_s(\mu^2_0)$, $\widetilde{\alpha}_s(\mu^2_0)$ are, respectively, 
$$\mu^2_0=7.86\,\gev^2,\; \alpha_s(\mu^2_0)=0.257\;, \widetilde{\alpha}_s(\mu^2_0)=0.415.$$
We see that $\alpha_s$ is small, thus justifying the use of 
perturbation theory. Moreover, $a/2\simeq(2.8\,\gev)^{-1}\ll\lambdav^{-1}$: 
the quarks move well away from the confinement region. Finally, 
the binding energy is also considerably larger than $\lambdav$. 
Thus we find our approximations justified {\sl a posteriori}.

 For $\bar{c}c$, and with 
$\lambdav(n_f=3,\,\hbox{three loops})=
0.338\pm0.037$ and the mass of the $J/\psi$ as an input now,
$$\eqalign{m_c=&1.936^{+0.059}_{-0.068}\,(\Lambdav)\;\mp0.014\,(\langle\alpha _sG^2\rangle)\;\cr
&^{-0.124}_{+0.106} \;(\hbox{varying}\; \mu^2\;{\rm by}\,25\%)\;\pm\;0.014\;
({\rm th.\;uncert.})\phantom{\quad(P)}\cr
\bar{m}_c(\bar{m}_c^2)=&
1.564^{+0.086}_{-0.035}\,(\Lambdav)\,\mp 0.013\,(\langle\alpha_sG^2\rangle)\;\cr
&^{-0.095}_{+0.119} \;(\hbox{varying}\; \mu^2\;{\rm by}\,25\%)\;\pm\;0.013\;
({\rm th.\;uncert.}),\cr}
\equn{(3.2)}$$
and $\mu^2_0=2.871\;\gev^2$ now. The errors, and the 
values of $\alpha_s$, $\widetilde{\alpha}_s$ 
increase correspondingly and it follows that 
the calculation is much less reliable than for the $\bar{b}b$ case, 
as the errors in (3.2) show.

The values of the $b$ quark masses  reported here, 
e.g., \equn{(3.1)}, 
are slightly larger than those one finds with the 
sum rule method (see for example, refs.~19). 
It is not clear to the author why this occurs. 
I suspect that the sum rule evaluations contain 
systematic uncertainties which are not under control; 
and indeed, the determinations are not very compatible one with 
another. Anyway, the discrepancies are not terribly large.

\booksubsection{3.2.  $n=2$ states}
For the states with $n=2$, the energy levels can
  be evaluated using the values found for $m_b$ and taking now  
$\mu=1/a$. However, since (as stated) 
radiative and nonperturbative corrections are large, the results are very 
sensitive to the value of $\mu$ chosen. For this reason it is more profitable 
to {\sl fit} $\mu$. This is the procedure followed in ref.~14, from where 
the following table for the mass splittings of the states shown is taken:
 
\medskip
\setbox1=\vbox{\offinterlineskip\hrule
\halign{
&\vrule#&\strut\hfil#\hfil&\quad\vrule\quad#&\strut\quad#\quad&\quad\vrule#&\strut\quad#\cr
 height2mm&\omit&&\omit&&\omit&\cr 
& \kern.5em States&&Theory&&Experiment\kern.3em& \cr
 height1mm&\omit&&\omit&&\omit&\cr
\noalign{\hrule} 
height1mm&\omit&&\omit&&\omit&\cr
&  $\phantom{\big{|}}2^3P_2-2^3P_1$&&$21\pm7$&&$21\pm1\;\mev$\phantom{l}& \cr
&  $\phantom{\big{|}}2^3P_1-2^3P_0$&&$29\pm9$&&$32\pm2\;\mev$\phantom{l}& \cr
&  $\phantom{\big{|}}2^3S_1-\overline{2^3P}$&&$181\pm60$&&$123\pm1\;\mev$\phantom{l}& \cr
&  $\phantom{\big{|}}2^3S_1-1^3S_1$&&$428\pm105$&&$563\pm0.4\;\mev$\phantom{l}& \cr
&  $\phantom{\big{|}}\overline{2^3P}-2^1P_1$&&$1.5\pm1$&&\kern3em--& \cr
 height1mm&\omit&&\omit&&\omit&\cr
\noalign{\hrule}}
\vskip.05cm}
\centerline{\box1}
\smallskip 
\noindent Here we use standard spectroscopic notation; 
the common, fitted value of $\mu$ is somewhat above $1\,\gev$, and the errors 
given are {\sl only}  
those generated by the errors 
in $\Lambdav$, $\langle \alpha_sG^2\rangle$ given above. 
The overall agreement of theory and 
experiment is noteworthy, particularly considering that the value of 
the single free parameter, $\mu$, is the {\sl same} for all states. 
However, the situation is less satisfactory than what a cursory glance to 
the Table might suggest: both perturbative and nonperturbative
 corrections are large, and 
the results are somewhat unstable.

The {\sl wave functions} 
for states with $n=2$ present such large errors that
 the calculation using the methods described up till now become meaningless for them. 

\booksubsection{3.3. Spin-dependent shifts and leptonic decay rates}
The evaluation of spin-dependent shifts, and decay rates follow 
patterns similar to those of the spin-independent 
energy shift evaluations. The expressions one finds are\ref{11,14},
$$\eqalign{M(V)-M(\eta)=m\,\dfrac{C_F^4\alpha_s(\mu^2)\widetilde{\alpha}_s(\mu^2)^3}{3}
\left[1+\delta_{\rm wf}+\delta_{\rm NP}\right]^2\cr
\times\left\{1+\left[\dfrac{\beta_0}{2}\left(\log\dfrac{a\mu}{2}-1\right)
+\tfrac{21}{4}\left(\log C_F\widetilde{\alpha}_s+1\right)+B\right]\dfrac{\alpha_s}{\pi}
+\tfrac{1\,161}{8\,704}\,
\dfrac{\pi\langle\alpha_s G^2\rangle}{m^4\widetilde{\alpha}_s^6}\right\};\cr}\equn{(3.3a)}$$
$$\eqalign{\Gammav(V\rightarrow e^+e^-)=\Gammav^{(0)}\times\,
\left[1+\delta_{\rm wf}+\delta_{\rm NP}\right]^2\,(1+\delta_{\rm rad}),\cr
\Gammav^{(0)}=2\left[\dfrac{Q_b\alpha_{\rm QED}}{M(V)}\right]^2
\left(mC_F\widetilde{\alpha}_s(\mu^2)\right)^3;\cr
\delta_{\rm rad}=-\dfrac{4C_F\alpha_s}{\pi};\;\delta_{\rm wf}=
\dfrac{3\beta_0}{4}\left(\log\dfrac{a\mu}{2}-\gammae\right)\dfrac{\alpha_s}{\pi};
\cr}\equn{(3.3b)}$$
$$\eqalign{
\delta_{\rm NP}=\tfrac{1}{2}
\left[\tfrac{270\,459}{108\,800}+\tfrac{1\,838\,781}{2\,890\,000}\right]
\dfrac{\pi\langle\alpha_s G^2\rangle}{m^4\widetilde{\alpha}_s^6}.\cr}$$
Here $V=\upsilonv,\;J/\psi$. 
The corrections  are fairly large, particularly the radiative
 correction\ref{20} $\delta_{\rm rad}$. Because of this the calculation is
 less reliable than what one would have expected for $\bar{b}b$, 
and fails completely for $\bar{c}c$. 
With the values of $m_b$ found before, one has the numerical results,
$$M(\Upsilonv)-M(\eta)=53.3\pm5.3\,(\Lambdav)\,\pm5.3\,
(\langle\alpha_sG^2\rangle)\,
\pm10\,(\mu^2=7.859\pm25\%)\equn{(3.4)}$$
and
$$\Gammav(\Upsilonv\rightarrow e^+e^-)=1.143\pm0.11\,(\Lambdav)\,
\pm0.11\,(\langle\alpha_s G^2\rangle)\
\pm0.24\,(\mu^2=7.859\pm25\%).\equn{(3.5)}$$

 Higher order NP corrections due to some higher dimensional operators 
 are also known for the decay rate (see ref. 16). They 
would produce a shift in the decay rate of $\sim0.11\;\kev$,  
 smaller than the contribution of $\langle\alpha_s G^2\rangle$
 or the  uncertainty caused by e.g. 
varying $\mu$ around $\mu_0$. We do {\sl not} include 
 either in the evaluation or the error estimate.

The calculated value for the decay is in reasonable agreement with the experimental figure,
$$\Gammav_{\rm exp.}(\Upsilonv\rightarrow e^+e^-)=1.320\pm0.04\,\kev.$$

\booksection{4. Heavy quarkonia at long 
distances. Connection \hb 
between the long and short distance regimes}
Here we consider bound states of heavy quarks at {\sl long} 
distances. This certainly includes $\bar{c}c$ with $n>1$ and  $\bar{b}b$ 
with $n>2$; $n=1$ for the first and $n=2$  
for the second are somewhat marginal.
As stated in the previous section, perturbative QCD supplemented with 
leading NP effects fails now; but, fortunately, and since the average velocity 
of bound states decreases with increasing $n$, we expect the dynamics to be governed by a 
potential: our task is to determine it. This has been considered by
 a number of people\ref{17,21,22}. Here we will follow the Dosch--Simonov method, in  
 the version of ref.~17, which will allow us to establish connection 
with the short distance analysis of the previous section. 

The potential, that we denote by $U(r)$, is 
expected to exhibit a number of features. First of all, it should 
behave as $\sigma r$ at long distances, as follows from e.g. the lattice calculations. 
 Secondly, it should 
contain a coulombic piece, so we write
$$U(r)=-\dfrac{\kappa}{r}+U_{\rm NP}(r),\equn{(4.1)}$$
and, at short distances, one identifies 
 $\kappa=C_F\alpha_s+\hbox{radiative corrections}$.

 To find this potential we   
consider the Green's function in terms of the Wilson loop, working directly in the 
nonrelativistic approximation, and for large time $T$:
 for a $\bar{q}q$ pair:
$$\eqalign{G(x,\bar{x};y,\bar{y})=&
\int{\cal D}z\, {\cal D}\bar{z}\,\ee^{-(K_0+\bar{K}_0)}\langle W(C)\rangle,\cr
\langle W(C)\rangle=&
\int{\cal D}B\ee^{\ii\int_0^T\dd t(L_{\rm int}+L_{\rm rad})}.\cr}\equn{(4.2)}$$
Note that 
we treat the quarks in the nonrelativistic 
quantum mechanical formalism, appropriate because of their 
nonrelativistic motion. Thus, $K_0,\bar{K}_0$ are time 
integrals of the kinetic energies 
(nonrelativistic lagrangians) of quark and antiquark, 
$$K_0=\dfrac{m}{2}\int_0^T\dd \dot{\bf z}(t)^2,\quad
\bar{K}_0=\dfrac{m}{2}\int_0^T\dd \dot{\bar{\bf z}}(t)^2.$$
However, the gluons are treated fully field-theoretically. So 
the radiation lagrangian is $L_{\rm rad}=-\tfrac{1}{4}\int\dd^3r\,G^2.$
The Wilson loop operator corresponds to the contour $C$
 enclosing the $q,\,\bar{q}$ paths from time 0 to 
time $T$. It should include path-ordered 
parallel transporters for the initial and final states,
 $\phiv(x,\bar{x}),\,\phiv(y,\bar{y})$ with e.g. in 
matrix notation
$$\phiv(x,\bar{x})={\rm P}\exp\ii g\int^x_{\bar{x}}\dd z_{\mu}\,t_aB^a_{\mu}(z)$$
which we do not write explicitly.

 To take into account 
the nonperturbative character of the interaction it is convenient 
to work in the background gauge formalism and write $B_{\mu}=b_{\mu}+a_{\mu}$ 
where the $a_{\mu}$ represent the quantum fluctuations and 
 $b_{\mu}$ is a background field. This is constructed such 
that the vacuum expectation value of the Wick ordered 
products of the $a_{\mu}$, and of the mixed $a_{\mu}$, $b_{\mu}$ 
products  vanish. Therefore, we may express the gluon
 correlator in terms of $b_{\mu}$ only:
$$\eqalign{\langle :G(x)G(y):\rangle\rightarrow\langle :G_b(x)G_b(y):\rangle,\cr
G_{b,\mu\nu}=\partial_{\mu}b_{\nu}-\partial_{\nu}b_{\mu}+gb_{\mu}\times b_{\nu}.}$$
Expanding in powers of the background field $b_{\mu}$ we may
 write the Wilson loop average 
as
$$\eqalign{\langle W(C)\rangle=&\int{\cal D}a{\rm P}\ee^{\int_C\dd z_{\mu}\,a_{\mu}}\cr
+&\left(\dfrac{\ii g}{2!}\right)^2
\int{\cal D}a\,\int_C\dd z_{\mu}\int\dd z'_{\nu}\,{\rm P}\phiv_a(z,z')
b_{\mu}(z){\rm P}\phiv_a(z',z) b_{\nu}(z')+\dots\cr
\equiv& W_0+W_2+\dots\cr}\equn{(4.3)}$$
and the transporter $\phiv_a$ is constructed with only the quantum field $a$.
For the first term, $W_0$, the cluster expansion 
gives
$$\eqalign{W_0=Z\exp\left(\varphi_2+\hbox{higher orders}\right),\cr
\varphi_2=\dfrac{C_Fg^2}{4\pi^2}\int^T_0\dd t\int^T_0\dd t'\,
\dfrac{1+\dot{\bf z}\dot{\bf z}'}{{\bf r}^2+(t-t')^2}\cr
=C_F\alpha_sr^{-1}\int_0^T\dd t+O(v^2),}$$
i.e., the coulombic piece of the potential. ($Z$ is a constant that, in particular, 
includes regularization).

We keep now only the next piece, $W_2$ and neglect the $W_n,\,n>2$.
 At {\sl short} distances, 
this is justified because higher $W_n$ involve higher powers 
of the background fields. 
Thus the ensuing corrections are suppressed,  on dimensional grounds, by powers of 
$1/m^n$. For long distances, and although 
arguments have been advanced for the dominance of the $W_2$, we really have a model, 
the so-called ``stochastic vacuum model". 

For details of the evaluation of this first nontrivial piece, $W_2$, 
we refer to the lectures by Simonov.\fnote{For technical reason, 
should be performed in euclidean space returning to Minkowski space at the 
end of the calculation. We omit these subtleties here.} It produces 
a correction to the Green's function, $\delta G$, which in the static approximation is 
$$\eqalign{\delta G=
-\tfrac{1}{24}\int\dd^3r\int\dd^3r'\int r_i\dd\beta\int r'_i\dd\beta'\cr
\times G^{(S)}_C(r(T),r)G_C^{(8)}(r,r')G_C^{(S)}(r',r(0)).\cr}$$
 $G_C^{(S,8)}$ are the singlet, octet coulombic 
Green's functions. The reason for the 
appearance of $G^{(8)}$ are similar to 
those for the appearance of $H^{(8)}$ in (2.9).

We may then take matrix elements between coulombic states, $|nl\rangle$, 
and identify the ensuing energy shifts from the relation
$$G=G^{(S)}_C+\delta G\simeqsub_{T\rightarrow \infty} G^{(S)}_C (1-T\delta E_{nl}).$$
We then find the basic equation\ref{17},
$$\eqalign{\delta E_{nl}=\tfrac{1}{16}\int\dfrac{\dd^3p\dd p_0}{(2\pi)^4}
\int\dd \beta\dd\beta'\widetilde{\Deltav}(p)
\sum\langle nl|r_i\ee^{\ii {\bf p}(\beta-1/2){\bf r}}|k(8)\rangle\cr
\times\dfrac{1}{E_k^{(8)}-E_n- p_0}\langle k(8)|r'_i\ee^{\ii {\bf p'}(\beta-1/2){\bf r'}}
|nl\rangle.\cr}\equn{(4.4)}$$
The states $|k(8)\rangle$ are eigenstates of the octet Hamiltonian,
 with energy $E_k^{(8)}$; 
the $E_n$ are the coulombic energies.\fnote{Here we 
subtract the rest energy, $2m$, 
from both $E^{(8)}$, $E_n$.}  Finally, $\widetilde{\Deltav}(p)$ is defined 
in terms of the correlators, being the Fourier 
transform of 
$$\Deltav(x)=D(x)+D_1(x)+x^2\partial^2 D_1(x)/\partial x^2$$
and 
$$\langle g^2:G_{0i}(x)G_{0j}(0):\rangle
=\tfrac{1}{12}
\left[\delta_{ij}D(x)+x_ix_j\dfrac{\partial^2D_1}{\partial x^2}\right].$$
We may write, using Lorentz invariance, $\deltav(x)=\deltav(x^2/T_g^2)$,
 with $T_g$ the so-called 
correlation time. This will play an important role in what follows. 

We have now two regimes. If $\mu_T\equiv T_g^{-1}\gg |E_n|$ the velocity 
tends to zero, and the nonlocality also tends to zero as compared with 
the quark rotation period (which in the coulombic approximation 
would be $1/|E_n|$). We can now neglect, in \equn{(4.4)}, 
both $E_n,\,E_k^{(8)}$ as 
compared to $p_0$ so, after some elaboration, 
 we obtain the energy shifts 
$$\delta E_{nl}\simeq \langle nl|U_{\rm NP}|nl\rangle$$
 with
$$\eqalign{U_{\rm NP}(r)=&\dfrac{2r}{36}
\Big\{\int_0^r\dd \lambda\int_0^{\infty}\dd \nu\, D(\lambda,\nu)\cr
+&\tfrac{1}{2}\int_0^r\dd\lambda^2\int_0^{\infty}\dd\nu\,
\left[-2D(\lambda,\nu)+D_1(\lambda,\nu)\right]\Big\},\cr
D(\lambda,\nu)\equiv&D(x_0^2,{\bf x}^2),\;{\rm etc}.\cr}\equn{(4.5)}$$
At large $r$, and as this equation shows, we 
find $U_{\rm NP}(r)\simeq\sigma r$. Here $\sigma$ can be related to $T_g$ and the 
gluon condensate if we assume a model for $\deltav$. 
So, if e.g. we take an exponential ansatz for $\Deltav(x)$, as in ref.~22, we find 
$$\mu_T=\dfrac{\pi}{3\sqrt{2}}\,\dfrac{\langle \alpha_s:G^2:\rangle}{\sigma^{\frac{1}{2}}}
\simeq0.32\,\gev.$$

For small $r$ the limit of the expression in 
\equn{(4.5)} leads to\ref{21}, 
$$U_{\rm NP}(r)\simeq c_0+c_1 r^2.\equn{(4.6)}$$
This is {\sl different} from the behaviour expected from the Leutwyler--Voloshin analysis 
which gave a behaviour $\sim r^3$; but one should 
understand that the present derivation holds for $r\rightarrow 0$ 
but still $T_g^{-1}\gg |E_n|$, i.e., 
in a situation other than that where the Leutwyler--Voloshin 
analysis is valid. 

It may be stated  
that the analysis based upon the potential (4.5) gives a very 
good description of heavy quarkonia states\ref{23}. 
Note, however, that the description is not the only one available 
on the market; others, based e.g. on relativistic 
corrections to an assumed linear potential are given in the 
papers in ref.~24.

We next get the matching between the 
two regimes\ref{17}. For this we turn to the opposite situation, viz., 
 $T_g^{-1}\ll |E_n|$. 
Now we may approximate $\Deltav(x)\sim {\rm constant}$, so that 
$\widetilde{\Deltav}(p)\sim\delta_4(p)$ and \equn{(4.4)} becomes
$$\delta_{\rm NP} E_{nl}=\dfrac{\pi\langle \alpha_s:G^2:\rangle}{18}
\langle nl| r_i\dfrac{1}{H^{(8)}-E_n+\mu_T}r_i|nl\rangle,\equn{(4.7)}$$
which coincides exactly with the results of the Leutwyler--Voloshin 
analysis\ref{2} in the limit $T_g\rightarrow \infty$ ($\mu_T\rightarrow 0$): 
cf. \equn{(2.9)}. In 
fact, \equn{(4.7)} allows us to estimate the finite 
size corrections to the  Leutwyler--Voloshin NP effects, which improves still the 
agreement between theory and experiment\ref{17}.

\brochuresubsection{5. Further discussion of nonperturbative effects:\hb 
Renormalons, and saturation}
In the previous sections we have shown how QCD can give a very satisfactory account 
of the heavy quarkonia spectra, particularly of the lowest lying states; 
an understanding based on perturbative calculations 
supplemented by NP ones, in particular those 
associated with the gluon condensate. Here we address 
some questions related to that.

First, one may inquire about the 
connection of {\sl renormalons} 
with nonperturbative effects. 
We return to the one-gluon exchange diagram, 
\fig~2. If we dress the gluon propagator with loops as in \fig~5 then the 
corresponding potential, in momentum space, is
$$\widetilde{V}(k)=\dfrac{-4\pi C_F}{k^2}\,
\dfrac{4\pi}{\beta_0\log(k^2/\Lambdav^2)},\equn{(5.1)}$$
and we have substituted the one-loop expression for $\alpha_s(k^2)$. 
The expression (5.1) is undefined for {\sl soft} gluons, with
 $k^2\simeq\Lambdav^2$. As follows from the general theory of 
singular functions, the ambiguity is of the form $c\delta(k^2-\Lambdav^2)$: upon 
Fourier transformation this produces 
an ambiguity in the $x$-space potential of 
$\delta V(r)=c[\sin \Lambdav r]/r.$ 
At short distances we may expand this in powers of $r$ and 
find\ref{25,26}
$$\delta V(r)\sim C_0+C_1r^2+\cdots\,.\equn{(5.2)}$$
The same result may be obtained with the more traditional method of Borel
 transforms.
 The behaviour in 
\equn{(5.2)} coincides with the short distance behaviour of
 the nonperturbative potential 
$U_{\rm NP}(r)$ as determined in ref.~21, and \equn{(4.6)} here.

\setbox9=\vbox{
\medskip
\setbox0=\vbox{\hsize 3truecm\captiontype\figurasc{\noindent figure 5.}{One-gluon
 exchange, dressed 
with loops.\vskip1cm
\phantom{x}}}
\setbox1=\vbox{\hsize 9truecm\hfil \epsfxsize=6.8truecm\epsfbox{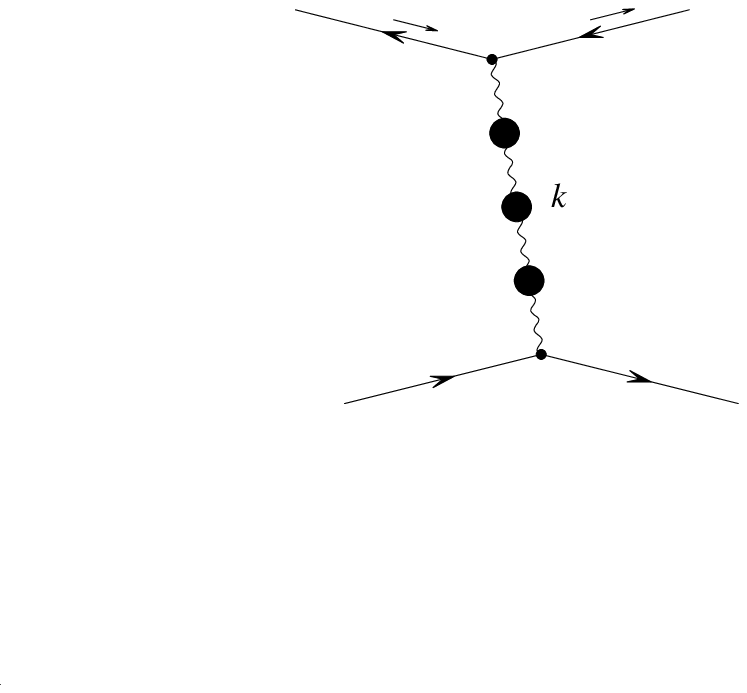}\hfil}
\line{\tightboxit{\box1}\hfil\box0}}
\box9

The situation just described applies for states $\bar{q}q$ at 
short distances; but not so short that zero frequency gluons cannot separate 
the bound state. If this last is the case, soft gluons do 
not resolve the $\bar{q}q$ pair and only {\sl see} 
a dipole. The basic diagram is no more that of \figs~2,~5, but that of \fig~6. 
The generated renormalon may then be seen\ref{26}
 to correspond to the contribution of the 
gluon condensate in the Leutwyler-Voloshin mechanism.

The matter of renormalons does not end here. If we calculate the 
renormalization of the mass to one loop, and dress the 
propagator with bubbles, one also finds a renormalon contribution 
to an ambiguity in the mass of\ref{27}
$$\delta_{\rm renormalon}m=\lambdav^2/m,$$
and there is another nonperturbative correction to the hamiltonian related 
to the arbitrariness in the origin of the energies. 
Indeed, we fix this origin by requiring the potential to vanish at 
infinity but, since the 
quarks are confined in a region of radius $R\sim\lambdav^{-1}$, ``infinity" 
is equivalent to $R$, hence we 
get an ambiguity of order $\lambdav\sim1/R$.
\topinsert{
\setbox9=\vbox{ 
\medskip
\setbox0=\vbox{\hsize 11.5truecm\captiontype\figurasc{\noindent figure 6.}{Emission 
and absorption of a soft gluon collectively by a $\bar{q}q$ pair.
\medskip}}
\setbox1=\vbox{\hsize 9truecm\hfil \epsfxsize=7truecm\epsfbox{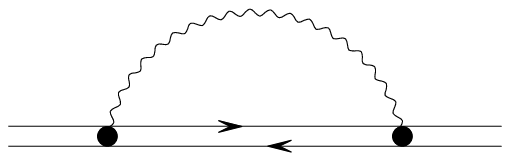}\hfil}
\centerline{\tightboxit{\box1}}
\medskip
\centerline{\box0}
\smallskip}
\box9}\endinsert

The situation, however, is less confused than 
what one might think. 
In fact, and at least in the case in which 
$m|v|$ is large compared to $\lambdav$, 
it can be shown that the linear and quadratic renormalons 
in the potential and pole mass cancel, 
leaving a $\lambdav^4r^3$ renormalon. 
This was to be expected on general grounds (see e.g., ref.~26); 
a formal proof may be found in ref.~28.

To make matters worse (or to improve them, depending on the viewpoint)
we will also consider 
the possibility of {\sl saturation}. We note that the 
ambiguities we have found  are 
associated with small momenta or,  equivalently,  long 
distances. However, at least the {\sl singularities} are clearly 
spurious. Indeed, not only the theory 
should be well defined but, because of 
confinement, long distances are never attained: the theory possesses an internal 
infrared cut-off 
of the order of the confinement radius, $R\sim\Lambdav^{-1}$. To try and 
implement it we  consider again the gluon propagator. To one loop 
it gets a correction involving the vacuum polarization tensor.
 Neglecting quarks this is, in $x$-space, 
given by an expression like 
$$\eqalign{\Piv^{aa'}_{\alpha\beta}(x,0)\sim& g^2f_{abc}f_{a'de}
\langle0|\int\dd^4y_1\,\dd^4y_2\,{\rm T}B_b^\alpha(y_1)\partial_\mu B_{c\alpha}(y_1)
B_d^\beta(y_2)\partial_\nu B_{e\beta}(y_2)|0\rangle\cr
+&\cdots\,.\cr}$$
We can take into account the {\sl long distance} interactions 
by introducing a string between the field products at finite distances. In 
 matrix notation for the gluonic fields, 
${\cal B}^\mu=t_aB_a^\mu$, this 
is implemented  by replacing
$${\cal B}^\alpha(y_1){\cal B}^\beta(y_2)\to
{\cal B}^\alpha(y_1){\rm P}\left(\exp\,\ii\int^{y_1}_{y_2}\dd z^{\mu}\,{\cal B}_\mu(z)\right)
{\cal B}^\beta(y_2).$$

The process may be described as ``filling the loop" (see \fig~7) by introducing 
all exchanges between the gluonic lines there. If we furthermore 
replace the perturbative vacuum $|0\rangle$ 
by the nonperturbative one $|{\rm vac}\rangle$, then a calculation
 similar to 
that made for the long distance potential for heavy quarks in \sect~4 
yields a dressed propagator

$$D^{\mu\nu}_{\rm dressed}(k)=
D^{(0)\mu\nu}(k)\dfrac{4\pi}{\beta_0\log(M^2+k^2)/\Lambdav^2},$$
and  $M^2$ is related to the gluon condensate at 
finite distances, $\langle G(x)G(0)\rangle_{\rm vac}$.

\topinsert{
\setbox0=\vbox{\hsize 8.2cm\hfil\epsfxsize 6cm\epsfbox{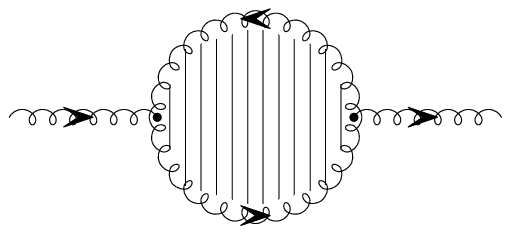}\hfil}
\setbox1=\vbox{\hsize 4cm\captiontype\figurasc{figure 7. }
 {``Filled in" gluon loop.}\hb
\vskip.8cm}
\setbox9=\vbox{\hsize 12.4cm
\line{\tightboxit{\box0}\hfil\box1}}
\centerline{{\box9}}
}\endinsert

This indicates a {\sl saturation} property of the coupling constant at 
small momenta (long distances);
 the calculation in fact suggests that, at small momenta, the expression for the 
running coupling constant should be modified according to
$$\alpha_s(k^2)=\dfrac{4\pi}{\beta_0\log k^2/\Lambdav^2}\to
\alpha^{\rm sat}_s(k^2)=\dfrac{4\pi}{\beta_0\log (k^2+M^2)/\Lambdav^2}.\equn{(5.3)}$$
It is certain that an expression such as (5.3) incorporates, to some extent, 
long distance properties of the QCD interaction. For example, if we take (5.3) 
with $M=\Lambdav$ in the tree level potential for heavy quarks, this 
becomes the {\sl Richardson potential}\ref{29}
$$\widetilde{V}^{(0)}({\bf k})=-\dfrac{4\pi C_F\alpha_s({\bf k}^2)}{{\bf k}^2}\to
\widetilde{V}^{(0),{\rm sat}}({\bf k})=
-\dfrac{16\pi^2C_F}{\beta_0{\bf k}^2\log (k^2+\Lambdav^2)/\Lambdav^2}.$$
When one has ${\bf k}^2\gg\Lambdav^2$, the short distance coulombic potential is, of course, 
recovered. For ${\bf k}^2\ll\Lambdav^2$, however,
$$\widetilde{V}^{(0),{\rm sat}}({\bf k})\simeqsub_{{\bf k}^2\ll\Lambdav^2}
\dfrac{16\pi^2C_F\Lambdav^2}{\beta_0{\bf k}^4},$$
whose Fourier transform gives
$${V}^{(0),{\rm sat}}(r)\simeqsub_{r\gg\Lambdav^{-1}}(\hbox{constant})\times r,$$
i.e., a linear potential. Indeed, a reasonably accurate description 
of spin-independent splittings in quarkonia states is
 obtained with such a potential.
 Likewise, use of (5.3) with $M=\Lambdav$ 
provides a surprisingly good description of small-$x$ deep inelastic 
scattering down to $Q^2\sim0$, as discussed in ref.~29; and these two cases 
are not unique.

In spite of these successes, it should nevertheless be obvious
 that (5.3) can only be of 
limited applicability. For example, consider the 
correlator of two currents  in the 
spacelike region, $\Piv(Q^2)$. We know that in some cases such as the 
correlators of vector or axial currents for massless 
quarks, or that of pseudoscalar ones, one has
$$\Piv(Q^2)\simeqsub_{Q^2\to\infty}\Piv_{\rm perturbative}
\left\{1+O(\langle\alpha_s G^2\rangle)Q^{-4}\right\},$$
whereas (5.3) would give a correction of order $M^2Q^{-2}$. 
The 
Richardson potential is also a good example of the limitations of the 
uses of saturation, in particular in connection with  
the extent to which saturation really does (or does not) 
represent a real, physical improvement, or merely the 
addition of a somewhat arbitrary new parameter. Indeed, the linear potential induced 
by saturation in the Richardson model is 
the {\sl fourth component} of  a Lorentz vector, 
while we know that the Wilson linear potential, as 
obtained, e. g., in the stochastic vacuum model or in lattice calculations, should be 
a Lorentz four-scalar: it thus follows that  
the linear potential obtained from saturation can be only of phenomenological
 use in some specific situations.

It is not easy to draw a clear morale from all of this. 
One can try to eschew the problem by expressing observables in terms of 
observables (for example, $\gammav(\upsilonv\to e^+e^-)$ in 
terms of $M(\upsilonv)$), hoping 
that this will reduce renormalon ambiguities, as some 
calculations seem to indicate\ref{27,30}. 
This is the viewpoint adopted in the papers in ref.~31. 
Another possible attitude is the following. 
It is very likely that the perturbative series in QCD are 
divergent; hence, 
different methods of summation lead to different results. 
This appears to be the case for renormalons or saturation resummations. 
It is the author's belief that only if the summation method 
is rooted on solid physics it is likely to represent an 
improvement; otherwise, estimates of nonperturbative effects 
become pure guesswork. In this respect, the method of taking 
into account the nonperturbative nature of the physical vacuum 
by considering the effect of nonzero values for the correlators 
stands some chance of being meaningful, as indeed phenomenological calculations 
seem to indicate.

\booksection{6.  Models}
\vskip-0.5cm
\booksubsection{6.1. The Constituent Quark Model}
We first discuss the constituent quark model. Here, we assume that 
the fact that quarks inside hadrons move through a 
medium made up of gluons and quark--antiquark pairs can, under certain circumstances, be 
represented by ascribing an effective mass, called the {\sl constituent} mass,  
even to light quarks. 
A possible way to connect this with a  QCD analysis might be the following.\fnote{Other 
mechanisms for the generation/interpretation of a constituent mass may 
be found in the lectures by Yu. Simonov, and in ref.~32.}
Consider the quark propagator, in the physical, nonperturbative vacuum that 
we denote by $|\rm vac\rangle$, 
$$S_{ij}(x)=\langle{\rm vac}|{\rm T}q_i(x)\bar{q}_j(0)|{\rm vac}\rangle;
$$
it is a gauge dependent object. We may define an invariant propagator by inserting 
a line integral. In matrix notation, but working 
in Minkowski space for now, we thus write 
an effective propagator as 
$$S_{ij}^{\rm eff}(x)=
-\dfrac{\delta_{ij}}{N_c}\langle{\rm vac}|{\rm T}\bar{q}(0)
{\rm P}\exp^{-\ii g\int_x^0\dd y^\mu\, {\cal B}_\mu(y)}q(x)|{\rm vac}\rangle.
\equn{(6.1)}$$
We can interpret $S_{ij}^{\rm eff}$ as the propagator describing a quark as it moves 
in the gluonic soup inside a hadron. In $p$-space,
$$S_{ij}^{\rm eff}(p)=\int\dd^4x\,\ee^{\ii p\cdot x}S_{ij}^{\rm eff}(x).$$
Evaluating the short distance limit with the OPE, we get the 
familiar lowest order expression 
$$S_{ij}^{\rm eff}(x)\simeqsub_{x\to 0}\delta_{ij}
\left\{\dfrac{-1}{4\pi^2}\slash{\partial}\dfrac{1}{x^2-\ii0}-
\dfrac{1}{4N_c}\langle\bar{q}q\rangle\right\};
\equn{(6.2)}$$
for simplicity we have taken the quark to be massless. 
At long distances we evaluate $S_{ij}^{\rm eff}(x)$ 
as follows. First, we go to Euclidean space. 
Then, and because we expect confinement (and thus that the 
interaction grows at long distances), we calculate for large coupling, $g\to\infty$. 
Finally, the quenched approximation is used.

Under these circumstances, the evaluation of $S_{ij}^{\rm eff}(x)$ is identical to 
that of the Wilson loop. Underlining Euclidean quantities, we then find
$$S_{ij}^{\rm eff}(\underline{x})\simsub_{\underline{x}\to\infty}
\delta_{ij}\ee^{-\sigma^{1/2}|\underline{x}|};$$
$\sigma$ is the string tension. In Minkowski space this becomes
$$S_{ij}^{\rm eff}(x)\simsub_{\underline{x}\to\infty}
\delta_{ij}\ee^{-\sigma^{1/2}\sqrt{-x^2}},
\equn{(6.3)}$$
an expression which is very appealing. According to it, the 
probability of a quark in the vacuum (inside a hadron) to propagate at a spacelike distance 
$r=\sqrt{-x^2}$ decreases exponentially when $r\gg \sigma^{-1/2}$; but the quark may  
move freely along a timelike or lightlike trajectory, where $\sqrt{-x^2}$ is 
pure imaginary or zero. 

A simple ansatz incorporating  short and long 
distance behaviour is
$$S_{ij}^{\rm eff}(x)=\delta_{ij}\left\{\dfrac{-1}{4\pi^2}\slash{\partial}\dfrac{1}{x^2-\ii0}-
\dfrac{1}{4N_c}\langle\bar{q}q\rangle\right\}
\ee^{-\sigma^{1/2}\sqrt{-x^2}}.
\equn{(6.5a)}$$
The corresponding $p$-space expression is then easily evaluated to be
$$S_{ij}^{\rm eff}(p)=\delta_{ij}
\left\{\dfrac{\ii}{\slash{p}}\left(1-\dfrac{\sigma^{1/2}}{(\sigma-p^2-\ii0)^{1/2}}\right)
-\dfrac{3\pi^2\ii \sigma^{1/2}\langle\bar{q}q\rangle}{N_c(K-p^2-\ii0)^{5/2}}\right\}.
\equn{(6.5b)}$$

The expression (6.5) for the propagator fulfills the Bricmont--Fr\"ohlich\ref{33} criterion
 for confinement and
 indeed exhibits many of the 
characteristics of the propagator for a particle with  
nonzero effective mass. Thus, $S_{ij}^{\rm eff}(p)$ presents a cut starting at $p^2=K$ 
and, what is more interesting, it behaves for $p\to0$ like the propagator for a massive particle:   
$$S_{ij}^{\rm eff}(p)\simeqsub_{p\to0}-
\dfrac{\ii\slash{p}}{2\sigma}+\dfrac{3\pi^2\ii\langle\bar{q}q\rangle}{N_c\sigma^2}
\simeq-\dfrac{\ii}{\mu_0},
\equn{(6.6)}$$
where the effective mass $\mu_0$ is
$$\mu_0=\dfrac{N_c\sigma^2}{-3\pi^2\langle\bar{q}q\rangle}.
\equn{(6.6)}$$
It is curious that in the last expression the quark condensate appears in the denominator.

The numerology  works reasonably well. With
 the value $\mu_0\simeq 320\,\mev$ obtained from phenomenological 
quark models, we can predict $\sigma^{1/2}\simeq 470\,\mev$, in reasonable agreement with 
lattice calculation results that give $\sigma^{1/2}\simeq 420\,\mev$.
These nice features should, however, 
not make one forget the shortcomings of our calculation here; (6.5) 
is to be considered  no more than a phenomenological expression. In fact, not 
only is the interpolation used  somewhat arbitrary, 
but, because the expression for the propagator only takes account 
of a certain class of gluon couplings, use of (6.5) into 
Feynman diagrams may lead to violations of gauge invariance. 
Because of this it is probably better not to ask too much of the model and 
take, simply, the consequence of the existence of a universal mass that
 represents the 
inertia acquired by quarks due to their having to drag in their motion the gluon--quark soup, 
and which adds to mechanical quark masses.
 Concentrating on light quarks, we then assume masses
$$m_u({\rm const})=m_u+\mu_0,\quad
m_d({\rm const})=m_d+\mu_0,\quad
m_s({\rm const})=m_s+\mu_0.\equn{(6.7)}$$

The presence of the mass $\mu_0$ breaks chiral invariance, 
and therefore pions and kaons (in particular) 
will be very poorly described by the constituent quark model: for these 
particles we have to use different methods. But one can use the 
constituent quark model to describe 
with success other 
hadrons ($\rho$, $K^*$, $\Sigmau$, $\Lambdau$, nucleons, $\Deltau$,\tdots).

To implement the interactions among quarks, we introduce two 
phenomenological potentials: 
a confining potential, linear 
in $r$,
$$U_{\rm conf}(r)=\lambda r,\quad
\lambda\sim \sigma,\equn{(6.8a)}$$
and a coulombic-type interaction,
$$U_{\rm Coul}(r)=\dfrac{-\kappa}{r},\equn{(6.8b)}$$
together with corresponding QCD-type hyperfine interactions. For quarks with 
indices $i,\,j$, we take
$$U_{\rm hyp}(r)=-\kappa\sum_{i\neq j}\dfrac{1}{m_im_j}\sum_at^a_it^a_j
\ybf{\sigma}_i\ybf{\sigma}_j,\equn{(6.8c)}$$
and $t_i,\,\ybf{\sigma}_i$ act on the wave function of quark $i$. 
$\kappa$ may be connected with the running coupling at, say, 
the reference momentum of 1 \gev:
$$\kappa\sim C_F\alpha_s(1\,\gev^2).$$
Because the model is in any case not terribly precise, one at times replaces the 
linear potential by a quadratic potential, which can be solved explicitly.

\booksubsection{6.2. The Bag Model}
In the bag model one neglects confinement, and replaces 
it by the boundary condition that quarks cannot 
leave a ``bag" with size $R\sim \lambdav^{-1}$. 
There are a number of variants of the bag: the 
quantum mechanical bag, the field theoretic bag, little bags, etc.\ref{34,35} 
The simplest, of course, is the quantum mechanical bag\ref{34}. 
There, as a first approximation,
 one considers quarks as free, subject only to the boundary 
condition that the wave function vanish for $r=R$. 
This bag is solved by writing the Dirac equation for free particles 
and imposing then the boundary condition. 
The simple model produces {\sl qualitatively} reasonable results 
for light quark states, with the exception of pionic and kaonic states: 
as was to be expected, because the presence of the bag breaks chiral invariance. 
It is also not clear to what extent the presence of the bag 
is a good simulation of the confinement 
mechanisms or, more generally, 
nonperturbative effects:  
in fact it is {\sl not} for heavy 
quarkonia. To see this, consider 
 a system of heavy quarks, $\bar{q}q$, inside a spherical bag of 
radius $R$. We take the interaction to be the coulombic one, 
$-C_F\alpha_s/r$, and impose the bag boundary condition. 

Let us denote by $m_r$ to the 
reduced quark mass, $m_r=m/2$ for quarkonium. For a state with energy $E$, 
define $n_E\equiv\sqrt{{\rm Ry}/(-E)}$, and 
the variable 
$\rho=2r/n_E a$, 
where $a=1/m_r C_F\alpha_s$ and 
${\rm Ry}=\tfrac{1}{2}m_r (C_F\alpha_s)^2.$ 
For states with $l=0$, the differential 
equation obeyed by the radial wave function $\psiv_E$ is 
$$\eqalign{P''(\rho)+&\left(\dfrac{2}{\rho}-1\right)P'(\rho)+
\dfrac{n_E-1}{\rho}P(\rho)=0,\cr
\psiv_E(\rho)=&({\rm Const.})\times\ee^{-\rho/2}P(\rho).\cr}$$
Moreover we have the bag boundary condition 
$P(L)=0$, 
where $L=2R/n_Ea$.

We are interested in the solution of this for 
$L\gg a$. To find it, we proceed as follows. The regular  
solution is proportional to  Kummer's function, 
$P(\rho)=M(1-n_E,2;\rho)$; 
the boundary condition then fixes $n_E$. To see this, consider 
the ground state. Since in the limit $L\to\infty$ we should 
recover the ordinary solution, 
with $n_E=1$, we write $n_E=1+\epsilon$ and work to 
lowest order in $\epsilon$. Expanding,
$$\eqalign{M(1-n_E,2;\rho)=&M(-\epsilon,2;\rho)
\simeq M(0,2;\rho)+\epsilon\dfrac{\partial}{a}M(a,2;\rho)\big|_{a=0}\cr
=&1+\dfrac{1}{\gammav(-\epsilon)}
\sum_{n=1}^\infty \dfrac{\gammav(n)}{n!\gammav(n+2)}\rho^n
\simeq1-\epsilon\eta(\rho),\cr
\eta(\rho)\equiv\sum_{n=1}^\infty\dfrac{\rho^n}{n (n+1)!}.\cr}$$
For this to have a zero at $\rho=L$  we must have
$$\epsilon^{-1}=\eta(L)=\sum_{n=1}^\infty\dfrac{L^n}{n (n+1)!}.$$
At large $L$, 
$\eta(L)\simeqsub_{L\to\infty}\ee^L/L^2$, 
hence
$\epsilon\simeq L^2\ee^{-L}=(2R/a)^2\ee^{-2R/a}.$
For the energy, therefore,
$$E=-{\rm Ry}+2m_r^3R^2(C_F\alpha_s)^2\ee^{-2R/a}.$$
In the case of quarkonium, the nonperturbative shift induced 
in the ground state  
by the presence of the bag of radius $R$ is thus 
$$\delta_{\rm NP}E=\tfrac{1}{4}m^3R^2(C_F\alpha_s)^2\ee^{-RmC_F\alpha_s},$$
totally different from what is found both at short distances 
(as caused by the gluon condensate, \equn{(2.10)}) or 
at long distances, as given by a linear potential $\sigma r$ 
that yields something proportional to $\sigma/m\alpha_s$.

\booksection{Appendix: Constants}
We collect here, for ease of reference, some of the constants that appear in the 
radiative corrections to  
 heavy quarkonium states. 
$$\eqalign{\beta_0=11-\tfrac{2}{3}n_f;\beta_1=102-\tfrac{38}{3}n_f\cr
\beta_2=\tfrac{2857}{2}-\tfrac{5033}{18}n_f+\tfrac{325}{54}n_f^2\cr}$$
$$\eqalign{a_1=&\dfrac{31C_A-20T_Fn_f}{36}\simeq 1.47;\;b_1=\dfrac{C_F-2C_A}{2}\simeq-2.33;\cr
a_2=&\tfrac{1}{16}
\Big\{\left[\tfrac{4343}{162}+4\pi^2-\tfrac{1}{4}\pi^4+\tfrac{22}{3}\zeta(3)\right]C_A^2\cr
&-\left[\tfrac{1798}{81}+\tfrac{56}{3}\zeta(3)\right]C_AT_Fn_f-
\left[\tfrac{55}{3}-16\zeta(3)\right]C_FT_Fn_f+\tfrac{400}{81}T_F^2n_f^2\Big\}
; \cr}$$
$$c_2^{(L)}=a_1\beta_0+\tfrac{1}{8}\beta_1+\tfrac{1}{2}\gammae\beta_0^2.$$

$$B=\tfrac{3}{2}(1-\log 2)T_F-\tfrac{5}{9}T_Fn_f+\dfrac{11C_A-9C_F}{18}. $$

$$\eqalign{N_1^{(n,l)} = {\psi(1 + l + n)-1 \over 2};\qquad 
N_0^{(n,l)} = {1 \over 4}\psi(1 + l + n)\big[\psi(1 + l + n) -2\big] \cr
+{n \over 2} \Bigg\{
{ (n-l-1)! \over (n+l)!}
\sum^{n-l-2}_{s=0} {(s + 2\,l + 1)! \over s!\,(s + l + 1 - n)^3 }\cr
 +
{ (n+l)! \over (n-l-1)!}
\sum^{\infty}_{s=n-l} {s! \over (s+2\,l+1)!\,(s+l+1-n)^3 } 
\Bigg\}.}
$$

We remark that the definitions xof  $a_2$ and $b_1$ have 
been swapped, in contrast to refs.~11,~13,~14, 
but in agreement with refs.~10.


\brochuresection{References}
{\item{1. }{A. De R\'ujula, H. D. Georgi and S. L. Glashow, Phys. Rev. {\bf D12} (1975) 147.}
\item{2. }{M. B. Voloshin, Nucl. Phys. {\bf B154} (1979) 365 and Sov. J. Nucl. Phys. 
{\bf 36} (1982) 143; H. Leutwyler, Phys. Lett. {\bf B98} (1981) 447.}
\item{3. }{S. N. Gupta and S. Radford, Phys. Rev. {\bf D24} (1981) 2309 and (E) 
{\bf D25} (1982) 3430; S. N. Gupta, S.~F.~Radford 
and W. W. Repko, {\sl ibid} {\bf D26} (1982) 3305.}
\item{4. }{See, e.g., C. Itzykson and J. B. Zuber, ``Quantum 
Field Theory", McGraw-Hill, New York, 1980 for the Bethe--Salpeter 
method, and W. E. Caswell and, for the method of effective lagrangians, G. P. Lepage, 
Phys. Lett. {\bf B167} (1986) 437 and N. Brambilla et al., hep-ph/9907240,
 from where other relevant references may be
 found.}
\item{5. }{F. J. Yndur\'ain, ``The Theory of Quark and Gluon Interactions",
 third edition, Springer, Heidelberg 1999.}
\item{6. }{A. Akhiezer and V. B. Berestetskii,  ``Quantum 
Electrodynamics", Wiley, New York, 1963; 
 V. B. Berestetskii, E. M. Lifshitz and L. P. Pitaievskii, 
``Relativistic Quantum Theory", Pergamon, London, 1971; 
F.~J.~Yndur\'ain, ``Relativistic Quantum Mechanics", Springer, Heidelberg 1996.}
\item{7. }{R. Coquereaux, Phys. Rev. {\bf D23} (1981) 1365; R. Tarrach, 
Nucl. Phys. {\bf B183} (1981) 384.}
\item{8. }{N. Gray et al., Z. Phys. {\bf C48} (1990) 673.}
\item{9. }{W. Fischler, Nucl. Phys. {\bf B129} (1977) 157; 
A. Billoire, Phys. Lett. {\bf B92} (1980) 343.}
\item{10. }{M. Peter, Phys. Rev. Lett. {\bf 78} (1997) 602; 
Y. Schr\"oder, DESY 98-191 (hep-ph/9812205), 1998.}
\item{11. }{S. Titard and F. J. Yndur\'ain, Phys. Rev. {\bf D49} (1994) 6007.}
\item{12. }{W. Buchm\"uller, Y. J. Ng and S.-H. H. Tye, Phys. Rev. {\bf D24} 
(1981) 3003.}
\item{13. }{A. Pineda and F. J. Yndur\'ain, Phys. Rev. {\bf D58} (1998) 094022, 
and hep-ph/9812371, 1998.}
\item{14. }{S. Titard and F. J. Yndur\'ain, Phys. Rev. {\bf D51} (1995) 6348.}
\item{15. }{A. Pineda,  Phys. Rev. {\bf D55} (1997) 407.}
\item{16. }{A. Pineda, Nucl. Phys, {\bf B494} (1997) 213.
The {\sl perturbative} corrections of order 
$\alpha_s^5\log\alpha_s$ have been recently  
calculated by N.~Brambilla et al., hep-ph/9910238; they  
are {\sl not} included in the computation. 
They would give a small shift in the energies; 
for, e.g., the $\upsilonv$  mass, and for 
our choice of $\mu=2/a$, of 
$$\delta E_{10}=-m[C_F+\tfrac{3}{2}C_A]C^4_F\alpha_s^5(\log\alpha_s)/\pi
$$
amounting to a decrease of the $b$ quark mass by 7.6 \mev.}
\item{17. }{Yu. A. Simonov, S. Titard and  F. J. Yndur\'ain, 
Phys. Lett. {\bf B354} (1995) 435.}
\item{18. }{For $\alpha_s$, 
see J.~Santiago and F. J. Yndur\'ain, hep-ph/9904344 (1999), 
in press in Nucl. Phys. B, and references therein. Recent values for the gluon 
condensate may be found in H. G. Dosch and S.~Narison, 
Phys. Lett. {\bf B417} (1998) 173 ;  F. J. Yndur\'ain, 
hep-ph/9903457, 1999, both using sum rules; M. D'Elia, A. Di Giacomo 
and E. Meggiolaro, Phys. Lett. {\bf B408} (1997) 315 and 
Phys. Rev. {\bf D59} (1999) 054503, with 
lattice calculations.}
\item{19. }{S. Narison, Phys. Lett. {\bf B341} (1994) 73 and 
Acta Phys. Pol., {\bf B26} (1995) 687;
 M. Jamin and A. Pich, Nucl. Phys. {\bf B507} (1997) 334;
 A. H. Hoang, Phys. Rev. {\bf D59} (1999) 014039 
and hep-ph/9905550; M.~Beneke and A.~Signer, hep-ph/9906475.}
\item{20. }{R. Barbieri et al., Phys. Lett. {\bf 57B} (1975) 455; {\sl ibid.} Nucl. 
Phys. {\bf B154} (1979) 535.}
\item{21. }{H. Dosch, Phys. Lett. {\bf B190} (1987) 177; Yu. A. Simonov, 
Nucl. Phys. {\bf B307} (1988) 512 and {\bf B324} (1989) 56; H. Dosch and 
 Yu. A. Simonov, Phys. Lett. {\bf B205} (1988) 339. See 
also Yu. A. Simonov, these Proceedings.}
\item{22. }{See, e.g., M. Campostrini, A. Di Giacomo and S. Olejnik, Z. Phys. {\bf C31} (1986) 577 
and work quoted there.}
\item{23. }{A. M. Badalian and V. P. Yurov, Yad. Fiz. {\bf 51} (1990) 1368; 
Phys. Rev. {\bf D42} (1990) 3138.}
\item{24. }{Other methods than that of 
the ``stochastic vacuum model" may be found in e.g., 
E. Eichten et al., Phys. Rev. {\bf D21} (1980) 203; 
N. Brambilla, P.~Consoli and G. M. Prosperi, Phys. Rev. {\bf D50} (1994) 5878; 
N. Brambilla, E.~Montaldi and G. M. Prosperi, {\sl ibid.} {\bf D54} (1996) 3506; 
N. Brambilla and A. Vairo, {\sl ibid.} {\bf D55} (1997) 3974.}
\item{25. }{U. Aglietti and Z. Ligeti, Phys. Lett. {\bf B364} (1995) 75.}
\item{26. }{R. Akhoury and V. I. Zakharov,  
Phys. Lett. {\bf B438} (1998) 165; F.~J.~Yndur\'ain, 
 Nucl. Phys. B (Proc. Suppl.) {\bf 64} (1998) 433.}
\item{27. }{M. Beneke, A. Signer and V. A. Smirnov, 
Phys. Rev. Lett. {\bf 80} (1998) 2535; 
A. Czarnecki and K.~Melnikov, Phys. Rev. Lett. {\bf 80} (1998) 2531.}
\item{28. }{A. Pineda, Ph. D. Thesis, Univ. Barcelona (1998); 
N. Brambilla et al., Phys. Rev. {\bf D60} (1999) 091502  and 
hep-ph/9907240.}
\item{29. }{J. L. Richardson, Phys. Lett. {\bf B82} (1979) 272. 
For an application to deep inelastic scattering, 
see K. Adel, F.~Barreiro and F. J. Yndur\'ain, Nucl. Phys.,
 {\bf B495} (1997) 221.}
\item{30. }{A. H. Hoang and T. Teubner, hep-ph/9801397 (1998); A. H. Hoang et al., 
Phys. Rev. {\bf D59} (1999) 014039.}
\item{31. }{A. H. Hoang , Z. Ligeti and A. V. Manohar, hep-ph/9811239 (1999).}
\item{32. }{H. Georgi and A. Manohar, Nucl. Phys. {\bf B234} (1984) 189; 
H. Georgi, ``Weak Interactions and Modern Particle Theory", 
Benjamin, 1984. For a recent 
paper in this spirit, from where references can be found, see 
A.~A.~Adrianov, D. Espriu and R. Tarrach, Nucl. Phys. {\bf B533} (1998) 429.}
\item{33. }{J. Bricmont and J. Fr\"olich, Phys. Lett. {\bf 122B} (1983) 73.}
\item{34. }{N. N. Bogoliubov, Ann. Inst. H. Poincar\'e, {\bf 8} (1967) 163.}
\item{35. }{See the reviews of K. Johnson, Acta Phys. Polonica, {\bf B6} (1975) 865; 
V.~Vento et al., Nucl. Phys. {\bf A345} (1980) 413.}
\item{}{}}

\bookendchapter
\bye